\documentclass[sigconf]{acmart}
\AtBeginDocument{%
  \providecommand\BibTeX{{%
    \normalfont B\kern-0.5em{\scshape i\kern-0.25em b}\kern-0.8em\TeX}}}
\usepackage{wrapfig}

\copyrightyear{2026}
\acmYear{2026}
\setcopyright{cc}
\setcctype{by}
\acmConference[DIS '26]{Designing Interactive Systems Conference}{June 13--17, 2026}{Singapore, Singapore}
\acmBooktitle{Designing Interactive Systems Conference (DIS '26), June 13--17, 2026, Singapore, Singapore}
\acmDOI{10.1145/3800645.3812978}
\acmISBN{979-8-4007-2563-0/2026/06}

\newcommand{\system}[0]{Prop-Chromeleon}

\usepackage{xcolor} 

\begin{document}

\title[\system{}]{\system{}: Adaptive Haptic Props in Mixed Reality through Generative Artificial Intelligence}



\author{Haoyu Wang}
\affiliation{
\institution{Imperial College London}
\institution{Royal College of Art}
\city{London}
\country{United Kingdom}
}
\email{haoyu.wang.work@gmail.com}

\author{Fengyuan Zhu}
\affiliation{%
 \institution{University of Toronto}
 \city{Toronto}
 \state{Ontario}
 \country{Canada}}
\email{fyzhu@dgp.toronto.edu}

\author{Bingjian Huang}
\affiliation{%
 \institution{University of Toronto}
 \city{Toronto}
 \state{Ontario}
 \country{Canada}}
\email{bingjian20@dgp.toronto.edu}

\author{Zhecheng Wang}
\orcid{0000-0003-4989-6971}
\affiliation{%
  \institution{University of Toronto}
  \city{Toronto}
  \state{Ontario}
  \country{Canada}
}
\email{zhecheng@cs.toronto.edu}

\author{Ludwig Sidenmark}
\affiliation{%
 \institution{University of Toronto}
 \city{Toronto}
 \state{Ontario}
 \country{Canada}}
 \email{lsidenmark@dgp.toronto.edu}


\begin{abstract}


Mixed Reality (MR) aims to blend digital and physical worlds, but the absence of haptic feedback often breaks visual-tactile consistency. We introduce \emph{\system{}}, a MR system based on generative artificial intelligence (AI) that dynamically transforms everyday objects into adaptive passive haptic props through user-provided text prompts. 
Our AI pipeline performs generation and anchoring of virtual assets that align with the shape of physical props, allowing us to study how virtual content generation behaves under geometric and prompt-based constraints. We evaluate \system{}'s effectiveness through a generation study using varied object shapes and user prompts, combining quantitative shape similarity metrics with qualitative prompt fidelity analysis. Our user study further showcases \system{}'s improvements in perceived realism, immersion, and enjoyment compared to static baselines. 
These results show that shape-aware generation can support both believable haptic interaction and creative engagement in MR.

\end{abstract}

\ccsdesc[500]{Human-centered computing~Interaction techniques}
\ccsdesc[500]{Human-centered computing~Virtual reality}
\ccsdesc[500]{Human-centered computing~User studies}

\keywords{Mixed Reality, Generative AI, Passive Haptics}

\begin{teaserfigure}
  \includegraphics[width=\textwidth]{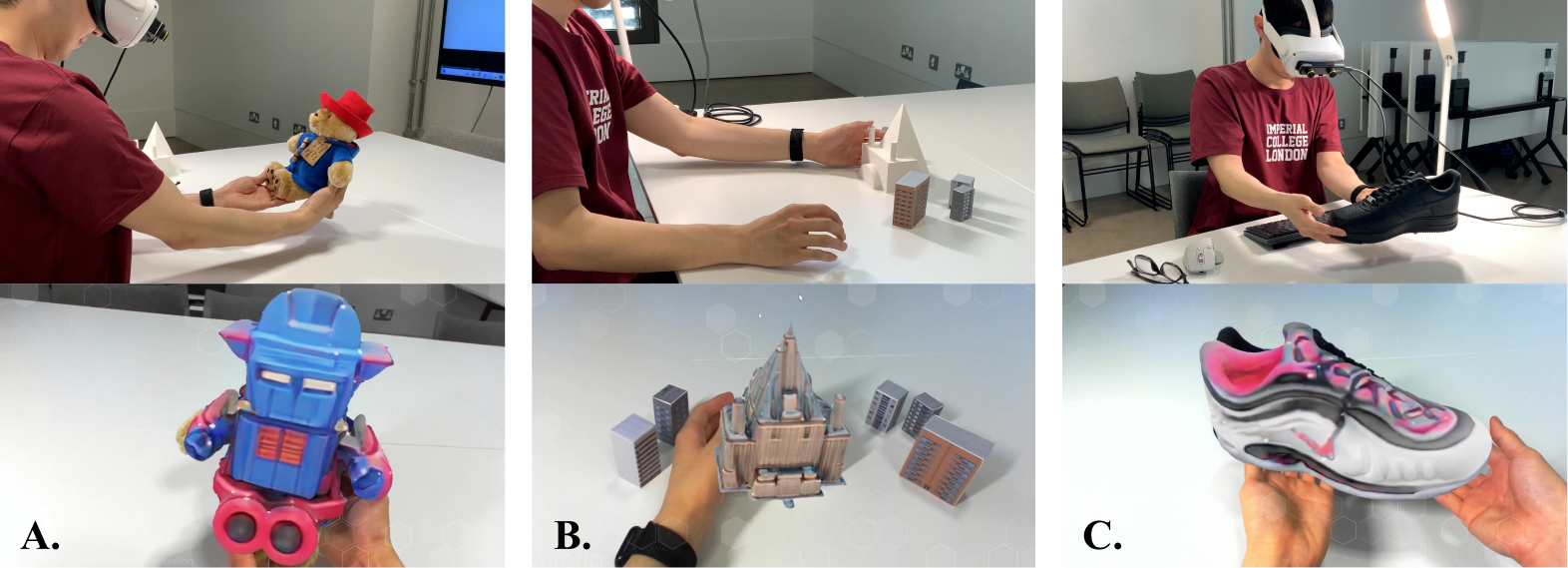}
  \caption{\system{} transforms physical objects into adaptive passive haptic props: A) transforming a Paddington bear toy into “a cute transformer toy”; B) transforming a 3D printed white building model into “an Empire State Building Architecture”;   and C) transforming a black Nike AF1 shoe into “a mix of Nike Air Foamposite One with Airmax 97 sneaker”.}
  \Description{
  A (top): A person dressed in a red t-shirt and wearing a VR headset holds a teddy bear dressed in a blue outfit with a red hat.
  A (bottom): A hand is shown holding a virtual, bear-shaped robot that appears to be rendered in a computer-generated style.
  B (top): A person's hands are seen arranging small-scale, simplistic white building models on a white surface.
  B (bottom): A hand is shown positioning models of buildings that have a rendered, digital appearance with a brown color scheme.
  C (top): A person in a red t-shirt, wearing a VR headset, is holding and examining a black sneaker closely.
  C (bottom): Two hands hold a sneaker that has a digitally-rendered appearance, featuring white and pink colors.
  }
  \label{fig:teaser}
\end{teaserfigure}


\maketitle

\section{Introduction} 
Mixed reality (MR) is designed to enable interaction with digital information that is always available and integrated into the physical environment~\cite{mixedreality2023}. The rise of consumer-grade MR devices (e.g., Apple Vision Pro~\cite{applevisionpro} and Meta Orion~\cite{meta_orion}) and increased interest from research have led to significant strides in integrating visual~\cite{Reipschlager2021Visualization, SpatialTouch, Prabha2024Enhanced} and auditory~\cite{DeepSpace, Tashev2017Head-related, Gupta2022Augmented/Mixed} content and information into our physical spaces. However, haptic experiences ~\cite{Hoffman1998tactile}, a critical component of how we interact with our surroundings, remain notably lacking in digital content, causing a disconnect between real-world and digital interactions that adds effort and reduces immersion. Enabling haptic experiences is especially challenging in MR due to the lack of control of the physical environment and the diversity of environments that MR is expected to operate in. These challenges require haptic solutions to not only be convincing in their haptic feedback but also adaptive to the current environment and needs of the user.

To enable compelling haptic experiences for virtual environments, researchers have proposed \emph{passive haptics}~\cite{Insko2001passive} that use everyday objects in the physical environment to mimic virtual objects, thus providing aligned tactile feedback. Compared to active haptic methods that require mechanical or electrical devices ~\cite{HapticDevicesTaxonomy}, passive haptics have the advantage of not requiring additional devices. In virtual reality (VR), passive haptics has seen use in a wide variety of scenarios and scales, ranging from leveraging everyday household objects~\cite{VRhapticsathome, RemixedReality} to world-scale environment transformations~\cite{VR3D_Recon, DreamWalker}. Because the user cannot see the physical object in VR, it is harder for the user to notice visual-tactile mismatch, thus achieving high immersion. In MR, however, building passive haptic systems is more challenging as even a small mismatch between the physical prop and the virtual object will be noticeable to the user. Researchers have attempted to address this issue with manual alignment~\cite{iTurk, VRhapticsathome}, predefined mapping rules ~\cite{annexingreality, HapticRetargeting}, or heavily crafted virtual asset libraries~\cite{ARchitect, annexingreality}. But these methods all require a significant amount of human labor, falling short in scalability and adaptability to diverse physical environments. Rather than manually selecting or mapping existing virtual assets to physical objects, we explore  generating virtual content that conforms to the physical world. This reframes passive haptics as a problem of conditional generation under physical constraints, where object geometry and user intent jointly shape the resulting interaction. To study this space, we developed a system that enables controlled investigation of generative passive haptics.

We present \emph{\system{}} for exploring this generative approach to passive haptics in MR. The key idea of \system{} is to build a generative artificial intelligence (GenAI) pipeline to adaptively create 3D models based on the user prompt and the shape and size of physical objects instead of relying on predefined physical-to-virtual mapping rules or static 3D asset libraries. \system{} takes text prompts as user input, captures the depth map of the physical object, and employs GenAI to create a corresponding 3D mesh with embedded color information that matches the user prompt, allowing the user or the application to control the generation. The generated mesh is then anchored and aligned with the physical object in real-time using AR authoring tools with 6-DoF (Degrees of Freedom) tracking, ensuring precise alignment during interaction. Unlike previous methods, \system{} does not require sourcing real-world proxies~\cite{annexingreality}, carrying compatible objects~\cite{DaiberEverydayProxy}, or following preset procedural steps~\cite{VRhapticsathome}. Instead, \system{} allows for scalable, flexible and automated passive haptic experiences in MR by leveraging GenAI.

We evaluated \system{} in two studies. First, in a technical evaluation, we tested \system{} on a wide range of object shapes and user prompts. We measured the generated shape disparity using established Chamfer and Hausdorff distances, and assessed overall generation results through qualitative coding. The results showed a 90\% successful generation rate, increasing with personalized prompts, and demonstrated \system{}'s ability to adaptively and accurately generate virtual objects according to the physical prop and the user prompt. In a second MR user study, we explored users' perceptions of \system{}'s performance and ability to provide passive haptic experiences with various physical objects and prompts. Our results showed that compared to an AI-generated baseline that does not consider the shape of the physical object, \system{} provides a significantly better perceived haptic experience and was preferred by the majority of study participants. In sum, our work contributes:
\begin{itemize}
     \item \emph{\system{}}, a scalable MR approach for creating adaptive passive haptics with everyday objects by generating virtual models with GenAI.
     \item Results from a technical evaluation showing how object shape and user prompts affect generation success and fidelity, and demonstrate that \system{} can handle a wide range of transformations across diverse shapes and prompts with a high success rate and strong geometric accuracy.
     \item Findings from a user study showing that users’ perceptions of passive haptics depend on the balance between physical alignment, semantic coherence, and user expectations, and that \system{} yields higher perceived realism, immersion, and preference than a baseline.
     
\end{itemize}
\section{Related Work}
Our work builds on previous work on passive haptics, virtual and physical integration, and GenAI.

\subsection{Passive Haptics for Extended Reality}
Haptic feedback is an essential part of any extended reality interaction, shown to increase the user's sense of immersion ~\cite{Insko2001passive, OpportunisticControls} and user performance ~\cite{Chandak2023Leveraging}. Compared to active haptic methods that require additional mechanical or electrical devices ~\cite{HapticDevicesTaxonomy}, passive haptics utilize existing objects as "physical props" that simulate the touch sensations of virtual objects ~\cite{Hinckley1994haptic, Hoffman1998tacticle}. Beyond providing tactile sensation, passive haptic props ground virtual objects in physical space, provide mechanical constraints that guide manipulation, and communicate object affordances such as solidity and scale, which together improve spatial perception, motor control, and realism~\cite{Hinckley1994haptic, Insko2001passive, Lederman2009}. Passive haptics has the key advantage of not requiring dedicated haptic devices that may be bulky, heavy, or expensive~\cite{Wang2019HapticReview}. Rather, it relies on ad-hoc matching and repurposing of physical objects to create haptic sensations.

In VR, researchers have explored various methods to repurpose physical objects and environments as virtual counterparts~\cite{bouzbib2021can}. \emph{VR Haptics at Home} by \citet{VRhapticsathome} addresses visual-tactile mismatches by manually aligning virtual objects with common household items, such as chairs and pillows, to serve as passive haptic props using pre-defined mapping rules defined by the user. \emph{Haptic Retargeting}~\cite{HapticRetargeting} warps the virtual space to redirect user hand movements while interacting with virtual objects, allowing a single physical object to serve as a passive haptic proxy for multiple virtual objects. In \emph{Sparse Haptic Proxy}, \citet{SparseHaptic} combined haptic retargeting with on-the-fly object mapping to create a set of geometric primitives that simulate
touch sensations in VR. Beyond object-level mappings, researchers also explored room-scale or even larger transformations. \emph{Smart Substitutional Reality}~\cite{SmartReality} converted a lab environment to a smart home environment with interactive functionality, while \citet{VR3D_Recon} used procedural generation rules to enable adaptability in dynamic environments. Finally, \emph{DreamWalker}~\cite{DreamWalker} went one step further to convert outdoor environments to VR scenes in real-time. Recent work has also explored reconstructing physical objects together with their interactivity in MR. For example, InteRecon captures both geometry and functional behavior of real-world artifacts, highlighting the importance of preserving interactivity alongside appearance in digital replicas~\cite{Li2025InteRecon}.
 In contrast to these works, we aim to integrate virtual objects into the user's physical environment through MR, requiring a higher level of integration between the physical and the virtual object.

\subsection{Virtual and Physical Integration}

Passive haptics in VR is more easily implemented as the user cannot see the physical object and thus has a higher tolerance for visual-tactile mismatch~\cite{SmartReality}. In contrast, implementing passive haptics in MR is much more challenging and requires a high level of integration between the virtual and physical environments. Findings from~\citet{Substitutionalreality} suggest that, while minor size differences are generally acceptable, shape mismatches significantly impact usability and believability. As the user can see both the virtual and the physical objects, any subtle mismatch between the pair will disrupt the immersive experience. To address this issue, \emph{Annexing Reality}~\cite{annexingreality} and Ubi-TOUCH~\cite{Jain2023UbiTouch} introduced a dynamic mapping approach. Given a virtual object, they scan the surrounding physical environment to identify the physical object with the most similar shape, and overlay the virtual object on top of it. Although effective, this approach is largely constrained by the available objects in the environment, as it is often difficult to find an object that shares the exact geometry. \emph{Remixed Reality}~\cite{RemixedReality} sought an alternative solution that actively changes the 3D rendering of the geometry and appearance of physical objects to fit the virtual interactions. However, this solution falls short in visual-tactile consistency when the user is heavily involved in hands-on experiences. 

Beyond haptic virtual and physical integration, researchers have investigated visual object substitution of physical objects in the wild~\cite{Kari2021TransforMR}, integration of virtual interfaces into the physical environment~\cite{Han2023BlendMR}, physical prototyping~\cite{Li2024AniCraft}, and avatar embodiment~\cite{Jiang2023HandAvatar}. These works demonstrate the promise of physical–virtual integration for creating meaningful experiences and fostering creativity, but they generally lack haptic feedback, or scalability and often limit integration to a specific type of object. More recent work has also explored integrating captured and generated content into collaborative MR environments. VirtualNexus augments 360° video with interactive environment cutouts and virtual replicas, enabling users to manipulate portions of real-world scenes~\cite{Huang2024VirtualNexus}. Similarly, Thing2Reality leverages GenAI to transform 2D content into interactive 3D objects in XR meetings, enabling rapid creation, sharing, and manipulation of artifacts~\cite{Hu2025ThingReality}.

In summary, previous virtual and physical integration solutions for passive haptics have significant constraints in manual setups, fixed mappings, and static virtual libraries. Our work is new in exploring an adaptive and scalable approach that adapts to user input with GenAI models to dynamically create and adapt virtual proxies to the characteristics of physical objects in real time. This allows users to generate virtual objects for experiences that are more aligned with the physical world.

\subsection{3D Content Creation with GenAI}
GenAI has seen significant advancements in delivering text, images, and video content~\cite{AIsurvey, Cao2022survey}. Text-to-image (T2I) diffusion models ~\cite{StableDiffusionPaper, Photorealistic, Ramesh2022HierarchicalTI} have been shown to be capable of producing high-quality images from text input, which have been made accessible to the public through applications such as Stable Diffusion~\cite{stabilityai2023}, ChatGPT-4o ~\cite{openai2024gpt4o}, and Midjourney~\cite{Midjourney}. ControlNet~\cite{controlnet} further enhances spatial control by incorporating sketches, soft edges, and depth as additional constraints for generation. In parallel, 3D reconstruction models have advanced significantly~\cite{NeRF, 3DGS, TripoGaussian, TripoSR, CRM}, allowing the transformation of 2D images into 3D representations such as meshes or point clouds. Combining models into multistage pipelines, such as text-to-image-to-3D, has gained traction in both academic research~\cite{ViewDiff, Instant3D, DreamStone} and the broader GenAI community~\cite{comfyui}.

Beyond static content generation, GenAI is emerging as a tool for solving problems and improving experiences in MR. Studies explore building conversational programming agents~\cite{LLMR, GenAIVR}, generating visual aesthetics for immersive environments~\cite{DeepSpace}, and improving creative design workflows with AI-assisted content generation~\cite{NeuralCanvas}. In this work, we leverage the scalability of GenAI to address the visual-tactile mismatch of passive haptics, while allowing users to retain control. More broadly, generative models are increasingly used as interactive components rather than purely content creation tools. In this work, we extend this perspective to passive haptics by treating it as a problem of generation under physical constraints.

\section{\system{}} \label{sec:system}
To enable on-demand passive haptic experiences across diverse everyday objects and user environments, we built \system{}, a wearable stereo MR system that dynamically transforms physical objects into haptic proxies. Unlike conventional passive haptic systems, \system{} leverages GenAI techniques to adaptively generate and overlay 3D models onto real-world objects in real-time, precisely aligning to their physical shape. Users initiate the transformation through text prompts, enabling intuitive hands-on interactions with everyday objects as if they had visually transformed into the prompted virtual assets. To realize this capability, we addressed two core challenges: (1) generating virtual models that match the shape and appearance of real-world objects by combining depth-based image generation and single-view 3D reconstruction with simple but intuitive, prompt-based user interaction; and (2) anchoring these generated models onto their corresponding physical objects within a wearable MR setup.


\begin{figure*}[t]
  \centering
  \includegraphics[width=\linewidth]{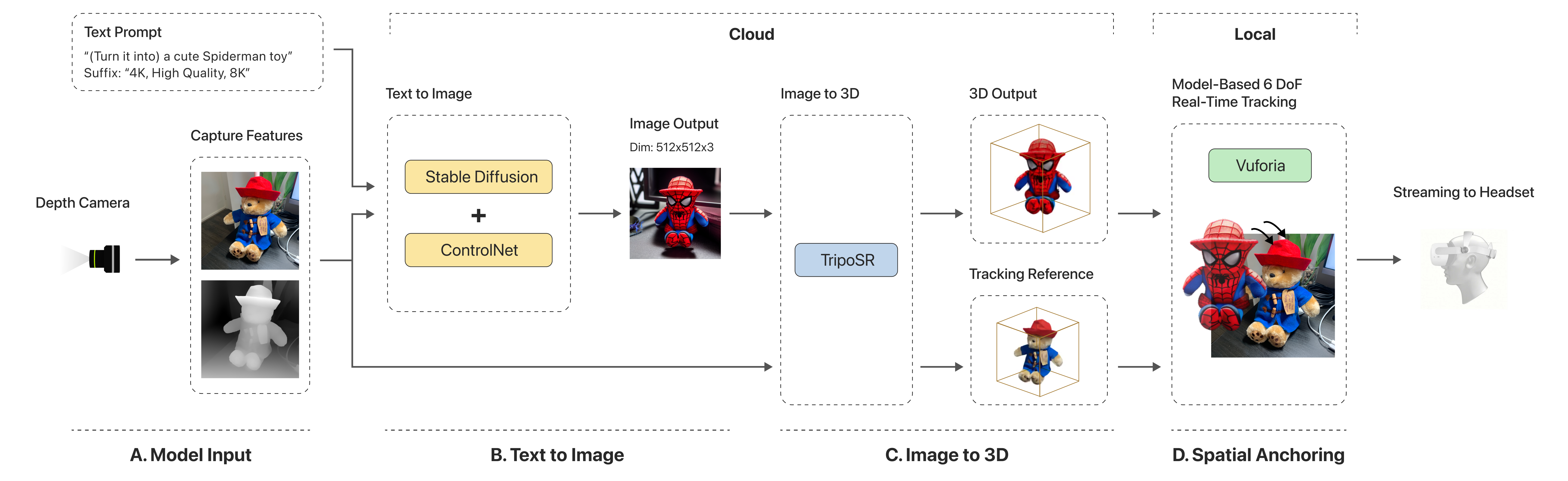}
  \caption{The \system{} processing pipeline combines depth and object tracking, and integrates the user's prompt by combining Stable Diffusion and ControlNet to generate a 3D object that is overlaid and aligned with the physical prop.}
  \Description{
  The image is a diagram labeled from left to right across four major steps, illustrating the process of converting a user text prompt into a 3D model displayed through a VR headset. 
  Model Input (A): The process begins with a "User's Text Prompt" asking to turn an object into "a cute Spiderman toy" with a specification for 4K, high-quality, 8K resolution. 
  Below this, a "Depth Camera" captures features of a teddy bear dressed in a blue suit and red hat, shown in two images: a colored photo and its depth map representation in gray.
  Text to Image (B): The features are processed in a cloud routine using "Stable Diffusion" combined with "ControlNet" to transform the captured image according to the text prompt.
  The "Image Output" shows a detailed Spiderman toy, retaining the teddy bear’s original position but transformed into a Spiderman theme.
  Image to 3D (C): The 2D image of the Spiderman toy is then converted into a 3D model using "TripoSR".
  Both the 3D model of Spiderman and the original teddy bear are shown within bounding boxes as "3D Output" and "Tracking Reference," respectively.
  Spatial Anchoring (D): The final step involves "Model-Based 6 DoF Real-Time Tracking" with "Vuforia" for spatial anchoring.
  The models are streamed to a VR headset, showing the user interacting with both the teddy bear and the Spiderman model in a virtual environment.
}
  \label{fig:pipeline}
\end{figure*}

\subsection{User Journey}
\begin{figure}[t]
  \centering
  \includegraphics[width=\linewidth]{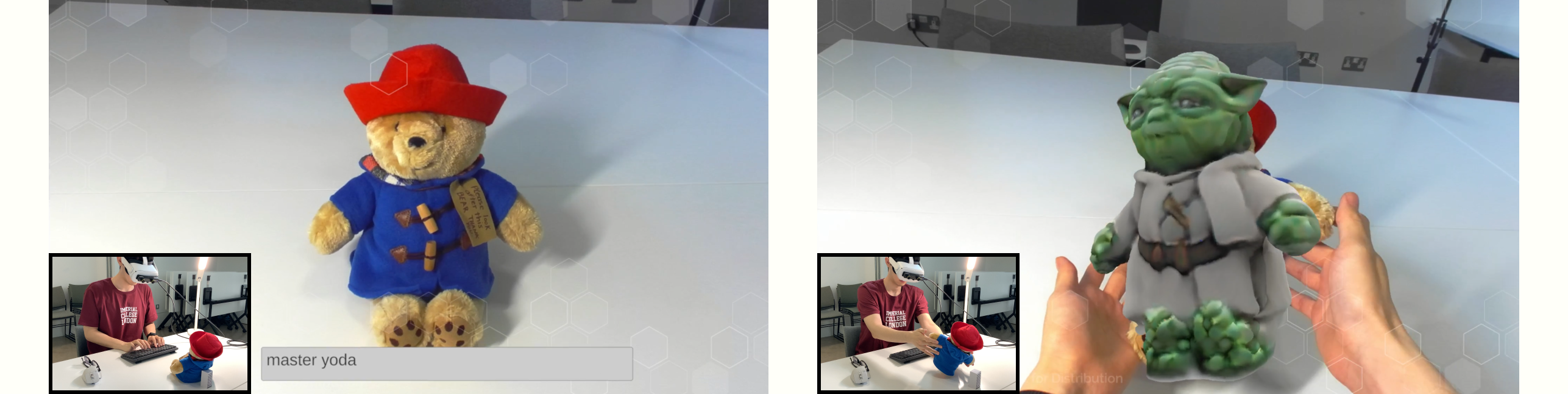}
  \caption{\system{} from the user's point of view. Left: The user inputs a "Master Yoda" text prompt to transform the Paddington bear. Right: \system{} generates a 3D model that matches the prompt and is overlaid on the physical prop for haptic interaction with the 3D model.}
  \Description{This image featuring two figures: The left half of the image shows the teddy bear, dressed in a blue suit and a red hat, positioned on a plain, white surface under bright lighting. Below the bear there's a input field floating on the screen. This part is also featuring a side view of a man wearing a VR headset typing on a keyboard, in front of his is the teddybear toy.
  The left half of the image shows the actual Master Yoda model is being held in the person’s hand, displaying detailed textures and color in green, white, and brown, indicating the figure's clothing and skin. The entire scene is overlaid with a hexagonal, futuristic grid pattern, enhancing the high-tech theme of the interaction. This part is also featuring a side view of the same man wearing the VR headset, and holding the same bear toy in his hand.
}
  \label{fig:userview}
\end{figure}

The user's experience with \system{} begins when they put on the MR headset and see a video passthrough view of their real-world environment. A minimal, floating input field appears just below their line of sight, prompting them to enter a text description of the object they wish to transform through text input or speech (\autoref{fig:userview}, left). For example, the user can type “Master Yoda” while visually centering a teddy bear positioned in front of them. Upon inputting the prompt, the teddy bear visually transforms into a virtual Master Yoda that precisely matches the bear's physical shape and remains consistently anchored even as the user moves around. This visual transformation allows the user to interact naturally with the physical object, picking up, holding and manipulating it, while visually perceiving and exploring the detailed virtual model corresponding to their prompt (\autoref{fig:userview}, right).

\subsection{System}
We developed \system{} using a mounted ZED Mini stereo camera on an Oculus Quest 2 headset. This hardware configuration is widely adopted in MR research for its combination of high-quality stereo passthrough and reliable depth-sensing capabilities~\cite{ChameleonControl, ARMixer, VisualNoiseCancellation, Hong2024, CookAR}.
To minimize perceived latency and mitigate rolling-shutter artifacts introduced by the ZED Mini, we set it to capture stereo video at a resolution of 720P at 60 frames per second (FPS). The resulting video feed is streamed directly to the Quest 2 and rendered in Unity (version 2022.3.20f1) at the same frame rate to maintain visual consistency. Both the ZED Mini camera and Quest 2 headset connect to a local Windows workstation (Windows 11, AMD 5950X CPU, NVIDIA RTX 3080Ti GPU, 32GB RAM) via cable. In Unity, we stream the camera feed onto a virtual curved display using the WebCamTexture class and external virtual camera software (e.g., OBS Studio~\cite{obsstudio}). The virtual display is rigidly positioned relative to the XR rig’s eye location, to create a low-latency MR passthrough experience that closely aligns with the user's physical viewpoint. We integrated a lightweight text-input interface into the Unity scene, featuring a simple floating input field labeled “Please type in your prompt.” Users input their transformation prompt using an external physical keyboard. 

\subsection{\system{} Pipeline}
\system{} explores a new generative pipeline beyond conventional reliance on static 3D asset libraries~\cite{ARchitect, annexingreality} and predefined mapping rules~\cite{annexingreality, HapticRetargeting}, which creates virtual content on demand, without manual alignment for greater scalability.

\subsubsection{Step A: Model Input}
The first stage of our software pipeline aims to precisely capture the physical object's shape and appearance as foundational input for generating a matching virtual model. When the user confirms their text prompt, the ZED Mini stereo camera simultaneously captures both a depth map and an RGB image of the object using the ZED API~\cite{zedAPI}.
The depth map, combined with the user's prompt, is then forwarded to the next stage (Step B, \autoref{fig:pipeline}) for spatially-conditioned image generation. Concurrently, the RGB image is passed directly to subsequent pipeline stages (Steps C and D), where it serves as the basis for generating the object's virtual appearance and constructing a reliable tracking reference. If depth capture hardware is unavailable, our system is extensible to use ControlNet's native depth estimation as an alternative.

\subsubsection{Step B: Text-to-Image}
In the second stage of our pipeline (\autoref{fig:pipeline}B), we perform depth-guided T2I generation to produce a 2D image matching the user's text prompt while precisely maintaining the captured shape of the physical object. This intermediate image forms the basis for subsequent 3D model reconstruction.

For T2I generation, we adopt Stable Diffusion 2.1~\cite{StableDiffusionPaper}, an open-source, diffusion-based generative model for its stable performance and extensive community validation. We specifically chose the version 1.4 checkpoint~\cite{sdv1.4} for its general-purpose versatility and compatibility. To integrate spatial constraints and preserve the object's shape, we employ ControlNet v1.1~\cite{controlnet}, selecting the depth-based conditioning method among available options, as it provides the most accurate representation of the object's geometric structure.
Due to the significant computational requirements of generative modeling, we remotely deploy the models from step B and step C using cloud services provided by HuggingFace Spaces~\cite{huggingfacespace} and Replicate~\cite{replicate_platform}. Our Unity-based application interacts with these cloud endpoints via real-time API calls.

\subsubsection{Step C: Image-to-3D}
In the third stage of our pipeline (\autoref{fig:pipeline}C), we reconstruct a 3D mesh from the depth-conditioned 2D image produced in Step B. Before reconstruction, background removal~\cite{rembg} is applied to isolate the object to prevent unwanted background artifacts from influencing the generated 3D model. Accurate background removal is critical, as residual elements can result in the shape mismatch or shift the resulting mesh's geometric center, adversely affecting tracking accuracy and visual alignment. In practice, we found that capturing the object against a clean, uncluttered background significantly enhances reconstruction quality.

For mesh generation, we use TripoSR~\cite{TripoSR}, an open-source neural model to generate a 3D model from a single view image. Compared to alternatives such as CRM~\cite{CRM} and TripoGaussian~\cite{TripoGaussian}, TripoSR strikes a balance between faster inference and output compatibility. Specifically, it produces meshes in the glTF format, which integrates seamlessly with Unity ~\cite{unitygltf} and is more suitable for runtime rendering than point cloud representations. The generated 3D mesh is then streamed back into the Unity environment at runtime.

\subsubsection{Step D: Spatial anchoring}
The final stage of our pipeline anchors the generated 3D model onto the corresponding real-world physical object. For accurate and stable tracking, we employ Vuforia Engine 10.22~\cite{vuforia}, a widely used MR tracking toolkit, via its Unity plugin. Specifically, we use Vuforia’s Model Target functionality~\cite{vuforia_model_targets}, which offers reliable six-degree-of-freedom (6-DoF) tracking using a pre-generated 3D model reference. We generate this tracking reference by reusing the same TripoSR-based single-view reconstruction method described previously, applied directly to the original RGB capture of the object (\autoref{fig:pipeline}, bottom branch of Step C). Although alternatives such as depth-based scanning~\cite{KinectFusion} or multiview neural reconstruction~\cite{MVR_MVD, MVR_MVLayoutNet, MVR_transformer} could offer higher accuracy, we chose TripoSR for its balance of inference speed, suitability to our single-view setup, simplicity, and tracking accuracy.

During runtime, once the physical object is detected, Vuforia instantiates a “Runtime Mesh” precisely aligned with the real-world object. We scale the prompt-generated 3D model (top branch of Step C, \autoref{fig:pipeline}) to match the bounding box dimensions of this “Runtime Mesh”, aligning their maximum extents along the x, y, and z axes. We guide users to initiate model generation from a frontal viewpoint
to ensure optimal visual alignment. 

The complete transformation process (depth capture, prompt processing, 3D reconstruction, and spatial anchoring) typically concludes within 20 seconds using our current cloud server configuration (single NVIDIA A10G GPU). Users can then seamlessly interact with the physical object, visually transformed into the virtual counterpart (\autoref{fig:userview}, right). We next use this system to evaluate how generative pipelines behave across object shapes and prompts, and how these properties translate to user experience.
\section{Study 1: Generation}
The \system{} system is designed to transform real-world objects into items prompted by the user while maintaining their original shape. To understand how generative pipelines behave under varying geometric and prompt constraints, we conducted a technical evaluation.
Our evaluation comprised two main components: a quantitative analysis of shape similarity and a qualitative assessment of the generative outcomes.


\begin{figure*}[t]
  \centering
  \includegraphics[width=\linewidth]{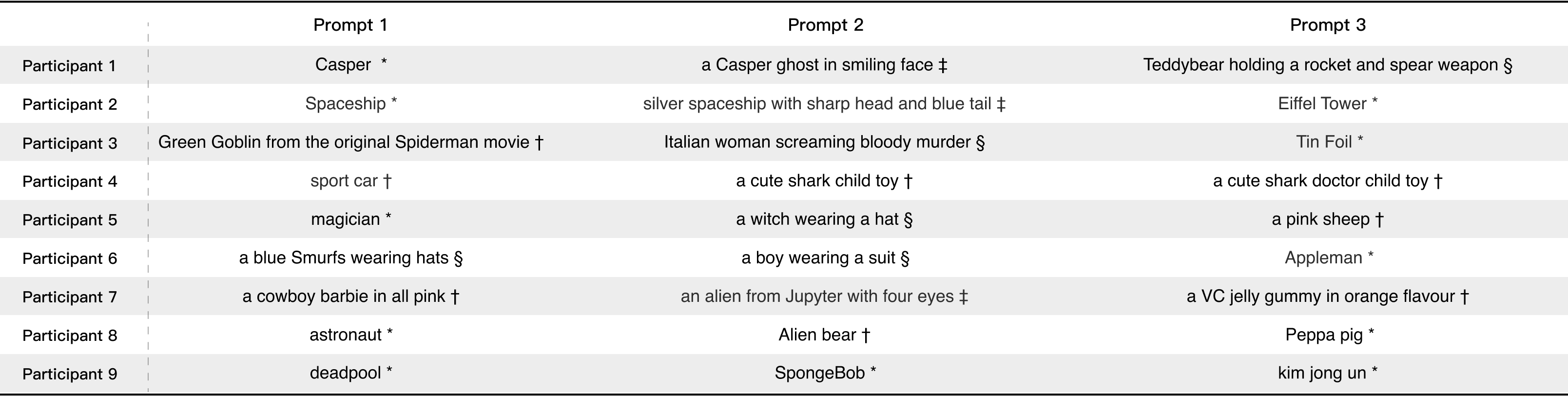}
  \caption{User-generated prompts from preliminary study. Symbols indicate the corresponding prop template specified in \autoref{sec:dataset}: * = `[Noun]', † = `[Adjective] [Noun]',  ‡ = `[Noun] with [Description]' and § = ‘[Noun] [Performing Action]’.}
  \Description{
  This image appears to be a structured table listing various user prompts given by nine different participants, categorized under three separate headings: Prompt 1, Prompt 2, and Prompt 3. Here’s a detailed breakdown of the information displayed in the table:
  Participants and Prompts
  Participant 1:
  Prompt 1: Casper *
  Prompt 2: A Casper ghost in smiling face ‡
  Prompt 3: Teddybear holding a rocket and spear weapon §
  
  Participant 2:
  Prompt 1: Spaceship *
  Prompt 2: Silver spaceship with sharp head and blue tail ‡
  Prompt 3: Eiffel Tower *
  
  Participant 3:
  Prompt 1: Green Goblin from the original Spiderman movie ‡
  Prompt 2: Italian woman screaming bloody murder §
  Prompt 3: Tin Foil *
  
  Participant 4:
  Prompt 1: Sport car †
  Prompt 2: A cute shark child toy †
  Prompt 3: A cute shark doctor child toy †
  
  Participant 5:
  Prompt 1: Magician *
  Prompt 2: A witch wearing a hat §
  Prompt 3: A pink sheep †
  
  Participant 6:
  Prompt 1: A blue Smurfs wearing hats §
  Prompt 2: A boy wearing a suit §
  Prompt 3: Appleman *
  
  Participant 7:
  Prompt 1: A cowboy barbie in all pink †
  Prompt 2: An alien from Jupyter with four eyes ‡
  Prompt 3: A VC jelly gummy in orange flavour †
  
  Participant 8:
  Prompt 1: Astronaut *
  Prompt 2: Alien bear †
  Prompt 3: Peppa pig *
  
  Participant 9:
  Prompt 1: Deadpool *
  Prompt 2: SpongeBob *
  Prompt 3: Kim Jong Un *
  
  Symbols Used: The symbols (*, †, ‡, §) could potentially represent categories or specifics about the prompts such as the source of the prompt, type of character, or some form of categorization pertinent to the study or survey from which these prompts were collected.
  }
  \label{fig:userprompts}
\end{figure*}

\subsection{Dataset Generation}\label{sec:dataset}
To comprehensively evaluate our system and ensure its relevance to real-user environments, our goal was to evaluate its behavior in a wide variety of shapes and prompts that users are likely to encounter in everyday scenarios. As such, we built a generation dataset consisting of a diverse range of shapes and prompts. First, we sampled 50 3D models with distinct shapes from ShapeNetCore v2~\cite{shapenet}, a large-scale 3D shape dataset, to cover a diverse range of geometric structures. We did not apply semantic constraints such as object category or expected real-world size when selecting 3D models, as our goal was to isolate the effect of geometry on generative alignment. Instead, we sampled objects to maximize diversity in shape, covering a broad range of geometric structures that could plausibly correspond to everyday handheld props (e.g., compact, manipulable objects). This allowed us to evaluate how different geometric profiles influence generation. 
Using Blender API~\cite{blenderAPI}, we generated depth maps for each model from the same standardized isometric viewpoint (35 degrees above ground, 30 degrees rotated horizontally) to replicate typical user perspectives.

For the prompts, we conducted a preliminary test involving nine participants (6 male, 3 female, average age=29.6, SD=11.76) to understand the types of prompts users naturally generate when interacting with \system{}. All participants were recruited from a local university campus, with backgrounds in design, fine art and engineering. Each participant was first introduced to the system, after which they were provided with a Paddington bear (\autoref{fig:userview}) as a reference object and asked to generate at least 3 different prompts. All prompts gathered are listed in \autoref{fig:userprompts}. We then categorized these prompts according to their syntactic and semantic characteristics, resulting in four identified prompt templates: 
\begin{enumerate}
    \item  ‘[Noun]’
    \item  ‘[Adjective] [Noun]’
    \item  ‘[Noun] with [Description]’
    \item  ‘[Noun] [Performing Action]’
\end{enumerate}

This ensured that the prompts we use in our analysis represented the prompts we could expect users to input. We then generated two sets of prompts based on these templates using ChatGPT 4 ~\cite{chatgpt}. The first set consisted of eight general prompts (two per template), designed without reference to object shape, and applied uniformly to all 50 models. The second set comprised object-specific prompts tailored to each model's geometry, and we used ChatGPT’s image capabilities~\cite{openai2024gpt4o} to assist this process. Again, we generated two prompts per structure, resulting in 8 prompts $\times$ 50 objects = 400 distinct prompts. We tested \system{} by applying general and object-specific prompts to the same set of 50 models, generating a total of 800 generations for evaluation. Please refer to our dataset repository\footnote{\url{https://github.com/harrywang7121/Prop-Chromeleon_Dataset}} for the selected models, prompts, and generated results.

\subsection{Shape Similarity Analysis}
Shape similarity is a crucial metric to evaluate, as visual-tactile mismatches can lead to negative user experiences. We evaluate shape similarity between the input geometry and the generated geometry from the pipeline using Hausdorff distance~\cite{Hausdorff} and Chamfer distance~\cite{chamfer}, both of which are well-established methods in visual computing to evaluate the geometry generation quality. These surface-based metrics are preferred over volumetric metrics like Intersection over Union (IoU) since many objects in popular datasets (e.g. ShapeNet) have hollow interiors. Hausdorff distance captures the worst-case geometric difference, while Chamfer distance offers a robust average-case analysis. To account for pose variations, we normalize both meshes into a unit sphere, apply random rotations, and perform rigid alignment before computing distances. This approach ensures that the results are not trapped in local minima, providing a more accurate and representative similarity assessment.

\begin{figure}
    \centering
    \includegraphics[width=\linewidth]{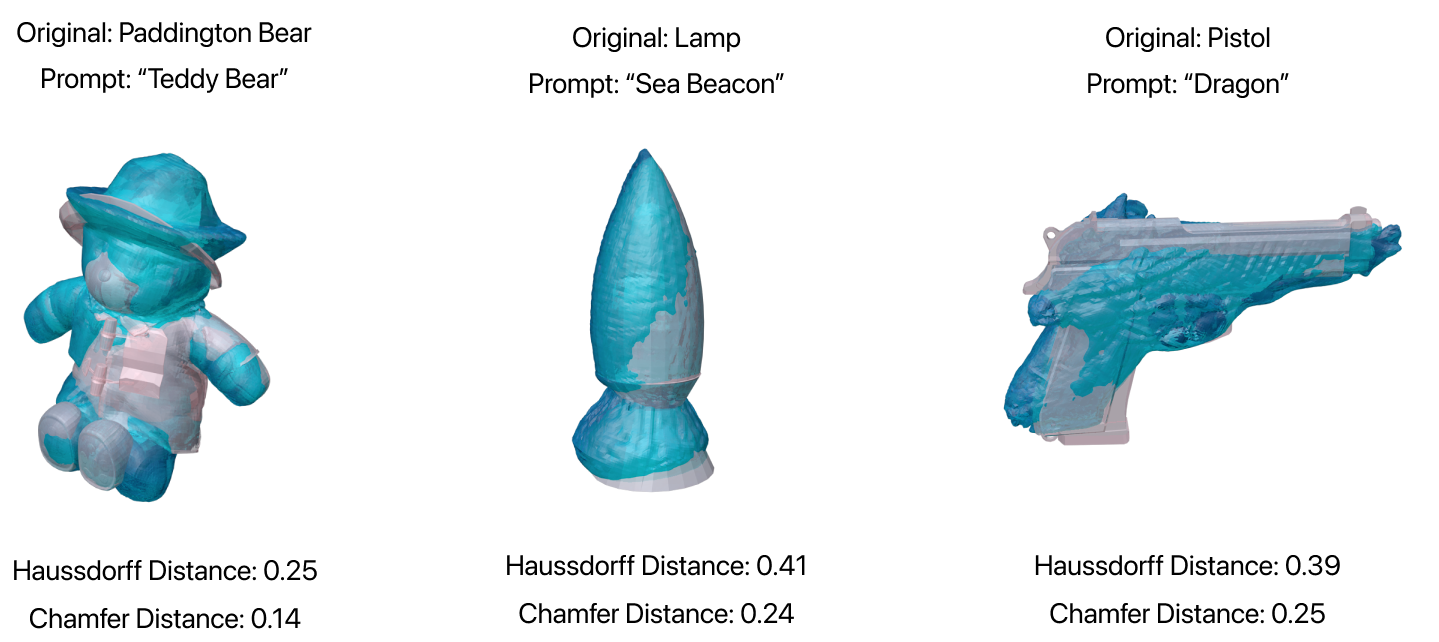}
    \caption{
    Comparison between generated meshes (blue) and original object shapes (white) produced by our pipeline. The Hausdorff and Chamfer distances quantify the geometric differences between the original and generated shapes.
    }
    \Description{
    This image showcases three different 3D models derived from prompts based on original objects, each annotated with specific metrics indicating the accuracy of the transformation:    
    Original Object: Paddington Bear
    Prompt: "Teddy Bear"
    Hausdorff Distance: 0.25
    Chamfer Distance: 0.14
    Description: The model shows a teddy bear wearing a blue coat and hat, lying down, with the transformation maintaining a high fidelity to the original Paddington Bear structure but simplified to a generic teddy bear form.
    Sea Beacon from Lamp
    
    Original Object: Lamp
    Prompt: "Sea Beacon"
    Hausdorff Distance: 0.41
    Chamfer Distance: 0.24
    Description: The lamp is transformed into a sea beacon resembling a lighthouse, maintaining cylindrical attributes of the lamp base but adding elements like a light chamber and windows to represent a sea beacon.
    Dragon from Pistol
    
    Original Object: Pistol
    Prompt: "Dragon"
    Hausdorff Distance: 0.39
    Chamfer Distance: 0.25
    Description: The pistol is morphed into the shape of a dragon, where the barrel represents the dragon's body and head, retaining some mechanical features of the pistol while incorporating organic dragon elements.
    }
    \label{fig:shape-comparison}
\end{figure}

\paragraph{Results}
We computed the Chamfer and Hausdorff distances over 50 models, each evaluated with 16 unique prompts from two conditions, resulting in a total of 800 input-generated geometry pairs. Across all geometry pairs we found a Chamfer distance of 0.322, and Hausdorff distance of 0.456. 
For consistent comparison, all shapes were normalized to a unit sphere with a radius of 1, where their size was scaled by the maximum displacement, and all objects were centered. In addition to numerical results, visual comparison (\autoref{fig:shape-comparison}) showed that the input and generated 3D models were almost identical, with only slight differences in fine details. The low Chamfer distance confirms minimal average deviations between the models, while the low Hausdorff distance indicates that even the largest differences were small and localized. These visual similarities indicate that the calculated distances were indeed very low, reflecting the high accuracy of the generated shapes.

\subsection{Qualitative Evaluation}
Due to the subjective nature and variety of the generated results, it is challenging to evaluate prompt fidelity for each group with a single success-or-failure metric. In our evaluation, we adopted a multi-dimensional approach and formulated three binary questions:
\begin{description}
    \item[Q1:] \textit{Does the model include the main elements from the prompt?}
    \item[Q2:] \textit{Does the model accurately represent specific features from the prompt, like color or texture?}
    \item[Q3:] \textit{Does the model reflect the intended theme of the prompt?}
\end{description}

We defined these three questions as we found that after reviewing multiple generations, the generated results of the prompts were expressed in three distinct ways. \emph{Q1} incorporates the main element into the designated shape. This assesses how well the generated object's shape fits into the original one. For example, a successful generation in this category could be a spherical object transformed into a crouching dragon that fits the shape of the object. A failed generation would be a ball with dragon scales but lacking other essential dragon characteristics, such as the head or eyes. \emph{Q2} represents the color or texture. For instance, while the dragon-scaled ball may fail in terms of shape, it would be a successful result in terms of texture. Some prompts, such as \texttt{``jewelry with gold and silver''}, express their features primarily through this category. \emph{Q3} represents the intended theme, which refers to the general perception of the result. Prompts like \texttt{``A talking cactus with a sombrero''} or \texttt{``futuristic vehicle''} can be interpreted in various ways when considering the first two categories. Also, sometimes the generated results for these prompts can be more abstract. Therefore, we included the intended theme to evaluate this aspect from a broader perspective.

We created a web interface for the analysis that presents all the necessary information for coding, including the prompt, the original object, the generated object, and a 360-degree rotating video of it. Each generation was coded by answering the three binary questions. A randomly selected subset of 112 generations was independently coded by three authors to ensure alignment between raters. We then used Fleiss' kappa to evaluate the alignment of the binary data between coders before reconciling disagreements. Overall, there was a very good alignment between the raters. Q1 had a very good alignment, $\kappa=.848$ (95\% CI, .741 to .955), $p<.001$, 105 of 112 aligned (93.75\%). Q2 had a very good alignment, $\kappa=.747$ (95\% CI, .640 to .854), $p<.001$, 111 of 112 aligned (99.11\%). Finally, Q3 also had a very good alignment $\kappa=.851$ (95\% CI, .744 to .958), $p<.001$, 104 of 112 aligned (92.86\%). After establishing a high alignment between coders, the rest of the dataset was rated by a single author.

\subsubsection{Results}
Overall, our results indicate that \system{} effectively generates 3D models that align with user prompts, particularly when prompts are customized to specific requirements. For Q1, which assesses whether the main elements from the prompt were included, custom prompts achieved a 95.5\% success rate, compared to 64.5\% for general prompts, with an overall rate of 80.0\%. In Q2, evaluating color and texture accuracy, the system performed consistently well, with a 97.0\% success rate for custom prompts and 90.3\% for general prompts (93.4\% overall). Q3, focusing on thematic accuracy, showed a 94.3\% success rate for custom prompts and 63.5\% for general prompts, with an overall rate of 78.8\%.

These results indicate that customized prompts generate better results in terms of key elements and thematic accuracy compared to general prompts. This improvement is due to the alignment between the original object and the target output, and the flexibility allowed by the prompt. For example, transforming a bucket into a mug yields more accurate results than transforming it into an owl, which requires a more constrained bird-like shape, limiting the system's ability to capture the necessary features. Similarly, prompts like "copper robot" achieve higher accuracy in both elements and theme than the owl prompt because robots allow for more flexible shapes. However, color and texture are less dependent on shape similarity, resulting in higher accuracy across both prompt types.

In conclusion, the \system{} system performs optimally when using custom prompts that consider the context of the original object, especially when there is flexibility in shape. Color and texture accuracy remain consistently high, while thematic alignment improves significantly with personalized input, highlighting the system's adaptability to user-specific needs.


\subsection{Summary}
In summary, our evaluation demonstrated \system{}'s ability to dynamically generate accurate and contextually relevant virtual models across diverse object shapes and prompts. The results also showed that generation was most successful when prompts were customized to match the object's form. Additionally, mesh similarity analysis using Chamfer and Hausdorff distances confirmed low geometric deviations between the input and generated models, validating the pipeline's effectiveness in maintaining shape accuracy. These results suggest that generation quality depends on both geometric compatibility and prompt specificity, highlighting key constraints of generative passive haptics.

\section{Study 2: User Tests}
To evaluate the effectiveness of \system{} in enhancing haptic experiences and achieving usable transformations, we conducted a user study focused on how shape alignment within \system{} improves haptic feedback, as well as collecting user feedback on the overall experience and transformation quality.

\subsection{Participants}
Twelve participants (7M, 5F) aged 23-54 (M=30.33, SD=10.44) took part in the study. Participants were asked to rate their experience with VR/AR/MR and GenAI on a scale of 1 (minimal experience) to 5 (extensive experience). The sample had moderate experience with VR/AR/MR (M=2.50, SD=1.38) and GenAI (M=2.67, SD=1.37).

\subsection{Conditions}
Two conditions were created to evaluate the impact of shape alignment on haptic feedback in \system{}. The key difference between the conditions was the alignment of the virtual model with the physical object.

\begin{figure}[t]
  \centering
  \includegraphics[width=\linewidth]{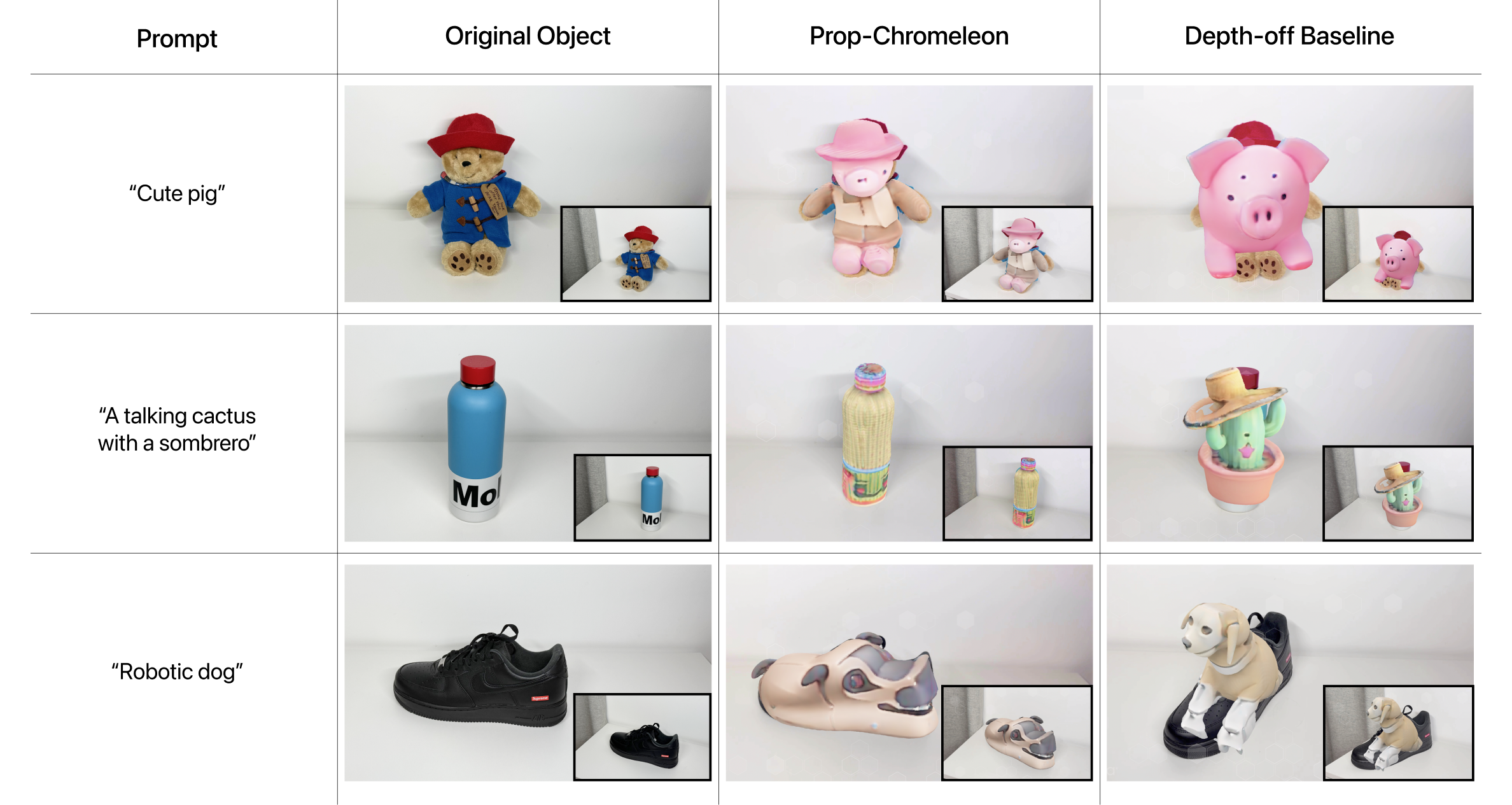}
  \caption{
  Prompts, original physical objects, Prop-Chromeleon generations, and baseline generations used for Task 1 in our user study.}
  \Description{
    The image showcase a visual display comparing different 3D model transformations based on specific prompts, categorized into three groups: the original object, a prop-transformed version, and a baseline model. Here’s a detailed description for each set shown:
    "Cute pig"
    Prompt: The transformation was to create a model resembling a "cute pig."
    Original Object: A teddy bear in a blue outfit and a red hat.
    Prop-Chromeleon: The bear is transformed with pig features, dressed in a light brown outfit with a pink pig face and a small hat.
    Baseline: A standard pink pig model with a simplistic design, sitting upright.
    "A talking cactus with a sombrero"    
    Prompt: The transformation was to create a model of a "talking cactus with a sombrero."
    Original Object: A plastic bottle with a blue cap.
    Prop-Chromeleon: The bottle is transformed to resemble a cactus wearing a colorful, woven sombrero.
    Baseline: A more traditional depiction of a cactus with a green body, two arms, a mouth, and wearing a straw sombrero.
    "Robotic dog"
    Prompt: The transformation was to create a "robotic dog."
    Original Object: A black sneaker.
    Prop-Chromeleon: The sneaker is modified with features resembling a futuristic, yellow robotic dog.
    Baseline: A more conventional robotic dog model, standing on four legs, with a design that combines elements of a sneaker and a dog.
  }
  \label{fig:conditiongroups}
\end{figure}

\textbf{\system{} Condition}: In this condition, our system was used as explained in \autoref{sec:system}, ensuring that the virtual and physical objects matched in shape (\autoref{fig:conditiongroups} middle).

\textbf{Baseline Condition}: In this condition, no depth-based shape alignment was applied. The generated virtual object was manually aligned with the physical prop to match its overall scale and orientation, as shown in ~\autoref{fig:conditiongroups}, right. This condition represents a scenario where the shape of the generated object is prioritized to match the prompt, rather than the physical object shape, allowing us to evaluate the added value of depth-based shape alignment.

Both conditions were tested across three scenes, each with a different physical prop: a Paddington Bear toy, a Nike Air Force One shoe, and a stainless steel bottle (\autoref{fig:conditiongroups}). The same virtual models were generated in both conditions using the following prompts: \texttt{``cute pig''} for the bear, \texttt{``yellow robotic dog''} for the shoe, and \texttt{``talking cactus with a sombrero''} for the bottle. The virtual models were anchored to their physical objects in Unity. 

\subsection{Tasks}

\begin{figure}[t]
  \centering
  \includegraphics[width=\linewidth]{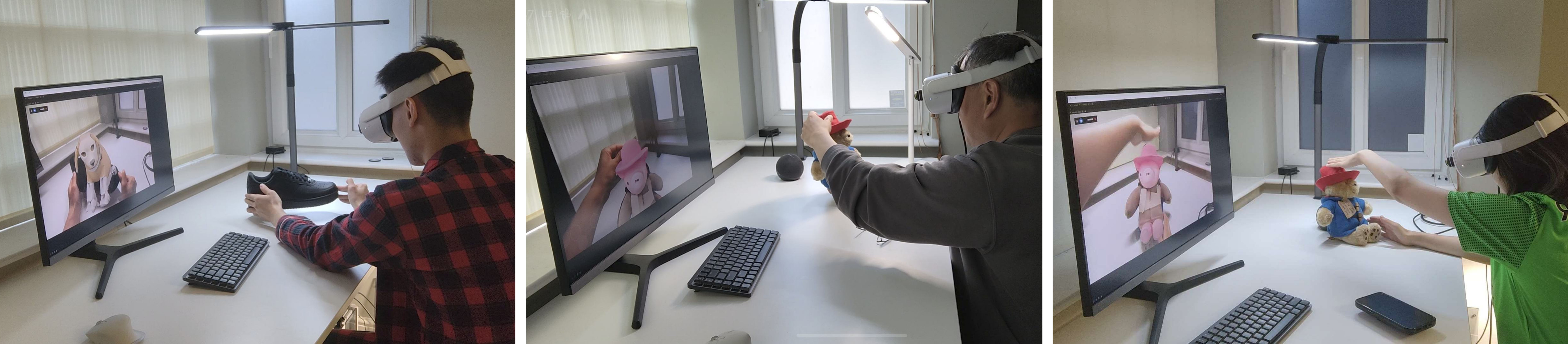}
  \caption{Participants interacting with different condition groups in the user study, Left: Participant moving the prop in the shoe-Robotic Dog scene, baseline condition; Middle: Participant touching the prop's hat brim in the Paddington Bear-Cute Pig scene, \system{} condition; Right: Participant engaging with prop in the Paddington Bear-Cute Pig scene, \system{} condition;}
  \Description{
    The image showcases a sequence of three photos depicting different individuals interacting with virtual reality technology in a modern office setting. Here’s a detailed description for each photo:
    Left Photo:
    Subject: A male individual wearing a VR headset and a red and black plaid shirt is seated at a desk.
    Activity: He is holding a black shoe with both hands, reflected on the monitor in front of him displaying a 3D virtual space with a model of a dog.
    Center Photo:
    Subject: Same as the first, viewed from behind.
    Activity: The individual interacts more closely with the virtual space, reaching out as if to touch or manipulate the object directly on the screen, showing a virtual model of a pink pig.
    Right Photo:
    Subject: A female individual in a green shirt, also wearing a VR headset, engaging with the virtual environment.
    Activity: She physically manipulates a teddy bear in a blue coat and red hat, directly correlating her actions to the virtual model shown on the screen in a similar pose.
  }
  \label{fig:userinteractions}
\end{figure}

\subsubsection{Task 1: System Comparison}

Task 1 was designed to compare the effectiveness of shape alignment in \system{} with a baseline condition. 
Participants interacted with three physical objects across both conditions: one where the virtual model was aligned with the physical object (using \system{}) and one where no alignment was provided (baseline). For each object, participants were asked to follow these instructions:
\begin{enumerate}
    \item Without touching it, observe the object from different angles. 
    \item Slowly hold the object with both hands, and move it to the other side of the table.
    \item Touch and feel the shape and outline of the object.
    \item Imagine it is a game, interact with the model in any way you prefer.
\end{enumerate}

\subsubsection{Task 2: \system{} Exploration}
In the second task, participants fully engaged with \system{} by transforming the same three physical objects from Task 1. Participants could choose any prompt for each object, and \system{} generated corresponding virtual models based on those prompts. After each transformation, they were asked to rate the generation result on "how well it met their expectations" on a scale of 1-5. Additionally, their behavior and prompt choices were observed and discussed to gather further insights into the user experience.

\begin{figure*}[t]
    \centering
    \includegraphics[width=\linewidth]{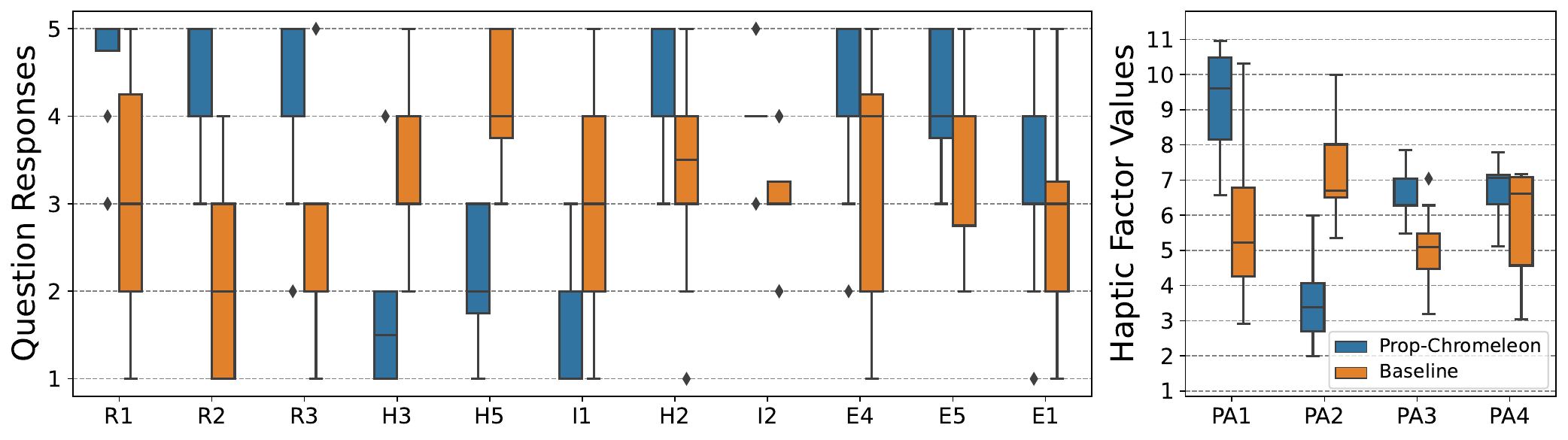}
    \caption{Questionnaire results for individual questions, and haptic factors. We found a significant difference for all questions and factors except  E5: \textit{``The haptic feedback reflects varying inputs and events''}, E1: \textit{``The haptic feedback all felt the same''},  and PA4: \textit{``Expressivity''}.}
    \Description{
    This image is a combined boxplot graph, split into two main sections that evaluate user responses and haptic feedback across different conditions in a study or testing scenario. Here's a detailed breakdown of the content displayed in each section:
    Section 1: Question Responses. This section shows boxplots representing the distribution of question responses. These are likely ratings provided by users based on a Likert scale (1 to 5) for various conditions labeled as R1, R2, R3, H3, H5, I1, H2, I2, E4, and E5.
    Colors: Two colors are used to represent different conditions or groups, likely "Prop-Chromeleon" and "Baseline".
    Data Points: Each boxplot displays the median (line inside the box), the interquartile range (height of the box), and range (whiskers). Outliers are shown as diamonds.
    Section 2: Haptic Factor Values. This part evaluates the haptic feedback factor, quantitatively measured on a scale, perhaps from 1 to 10, under four different conditions labeled as PA1, PA2, PA3, and PA4.
    Colors: Similar to the first section, two colors distinguish between "Prop-Chromeleon" and "Baseline".
    Data Points: Displays the distribution features similar to the first section, including the median, interquartile range, range, and outliers.
    }
    \label{fig:questionnaire-results}
\end{figure*}

\subsection{Procedure}
Before starting the study, participants signed a consent form and answered a demographic questionnaire. Each participant was then introduced to the concept of VR haptics, which included an explanation of the tactile sensations experienced when interacting with virtual objects in VR. Participants then put on the headset equipped with the attached depth camera system and were given adequate time to adjust it for comfort. Following this, they received a tutorial on using \system{}. Once participants were comfortable with the system, they were guided to proceed to the two tasks described, which were designed to evaluate \system{}’s effectiveness in enhancing haptic experiences. All participants began with Task 1, as shown in \autoref{fig:userinteractions}. To balance the study, half of the participants started with the \system{} condition, while the other half began with the baseline condition. After interacting with all three objects under one condition, participants completed a questionnaire based on the multidimensional Haptic Experience (HX) scale~\cite{HXscale}. The HX scale is an 11-item questionnaire designed to measure four key dimensions of haptic interaction: PA1) Realism (R1: \textit{``The haptic feedback was realistic''}, R2: \textit{``The haptic feedback was believable''}, R3: \textit{``The haptic feedback was convincing''}); PA2) Harmony (H3: \textit{``The haptic feedback felt disconnected from the rest of the experience''}, H5: \textit{``The haptic feedback felt out of place''}, I1: \textit{``The haptic feedback distracted me from the task''}); PA3) Involvement (H2: \textit{``I like having the haptic feedback as part of the experience''}, I2: \textit{``I felt engaged with the system due to the haptic feedback''}); and PA4) Expressivity (E4: `\textit{`The haptic feedback changes depending on how things change in the system''}, E5: \textit{``The haptic feedback reflects varying inputs and events''}, E1: \textit{``The haptic feedback all felt the same''}). After completing both conditions, participants engaged in a semi-structured interview where they discussed their ratings, system preferences, and how various factors influenced their experience. For Task 2, participants engaged in another semi-structured interview after the rating process, providing additional feedback on their experiences with \system{}, which we used to identify patterns in how participants interacted with the transformed objects and get a deeper understanding of user preferences and engagement.

\subsection{Results}

\newcommand{\wilcox}[4]{$z$$=$$#1$, $p$$#2$$#3$, $r$$=$$#4$}

We evaluated \system{}'s effectiveness and usability through two tasks. Overall, both tasks were conducted successfully. A combination of quantitative results and insights from post-study interviews led to several key findings.

\subsubsection{Questionnaires and Preferences}
When asked about their preferred system, eleven out of twelve participants stated that they preferred \system{}, while one participant mentioned the baseline. These results were further highlighted by the HX questionnaire results (\autoref{fig:questionnaire-results}), which were generally in favour of \system{}. The participants perceived that the haptic feedback for \system{} was significantly more \textit{``realistic''} (R1: \wilcox{2.59}{=}{.010}{.75}), \textit{``believable''} (R2: \wilcox{3.10}{=}{.002}{.89}), and \textit{``convincing''} (R3: \wilcox{2.87}{=}{.004}{.82}) than the baseline. These results were further confirmed by the significant difference for the combined \textit{``Realism''} factor (PA1: \wilcox{3.06}{=}{.002}{.88}).

The participants also indicated that the haptic feedback for the baseline was significantly more \textit{``disconnected from the rest of the experience''} (H3: \wilcox{3.13}{=}{.002}{.90}), \textit{``felt out of place''} (H5: \wilcox{2.98}{=}{.003}{.86}), and \textit{``distracting''} (I1: \wilcox{2.70}{=}{.007}{.77}) than \system{}. These results were further confirmed by the significant difference in the \textit{``Harmony''} factor (PA2: \wilcox{3.06}{=}{.002}{.88}). Results also indicated that for \system{}, participants significantly more \textit{``liked having the haptic feedback part of the experience''} (H2: \wilcox{2.88}{=}{.004}{.83}) and felt more \textit{``engaged with the system due to the haptic feedback''} (I2: \wilcox{2.76}{=}{.006}{.80}). These results were again confirmed by the combined \textit{``Involvement''} factor (PA3: \wilcox{3.06}{=}{.002}{.88}).

\textit{``Expressivity''} was the only factor for which we found no significant differences between \system{} and the baseline (PA4: \wilcox{1.68}{=}{.092}{.48}). Although the results showed that the participants found \textit{``the haptic feedback changes based on changes in the system''} (E4: \wilcox{2.07}{=}{.038}{.60}) significantly more for \system{}. We found no significant differences when asked whether each system's haptic feedback \textit{``all felt the same''} (E1: \wilcox{0.96}{=}{.336}{.28}), or whether \textit{``the haptic feedback reflects varying inputs and events''} (E5: \wilcox{1.55}{=}{.121}{.45}). These results are not surprising, since we only evaluated passive haptic systems where the haptic experience was static based on the shape of the physical object.

\subsubsection{Interview Results}

\begin{figure}[t]
  \centering
  \includegraphics[width=\linewidth]{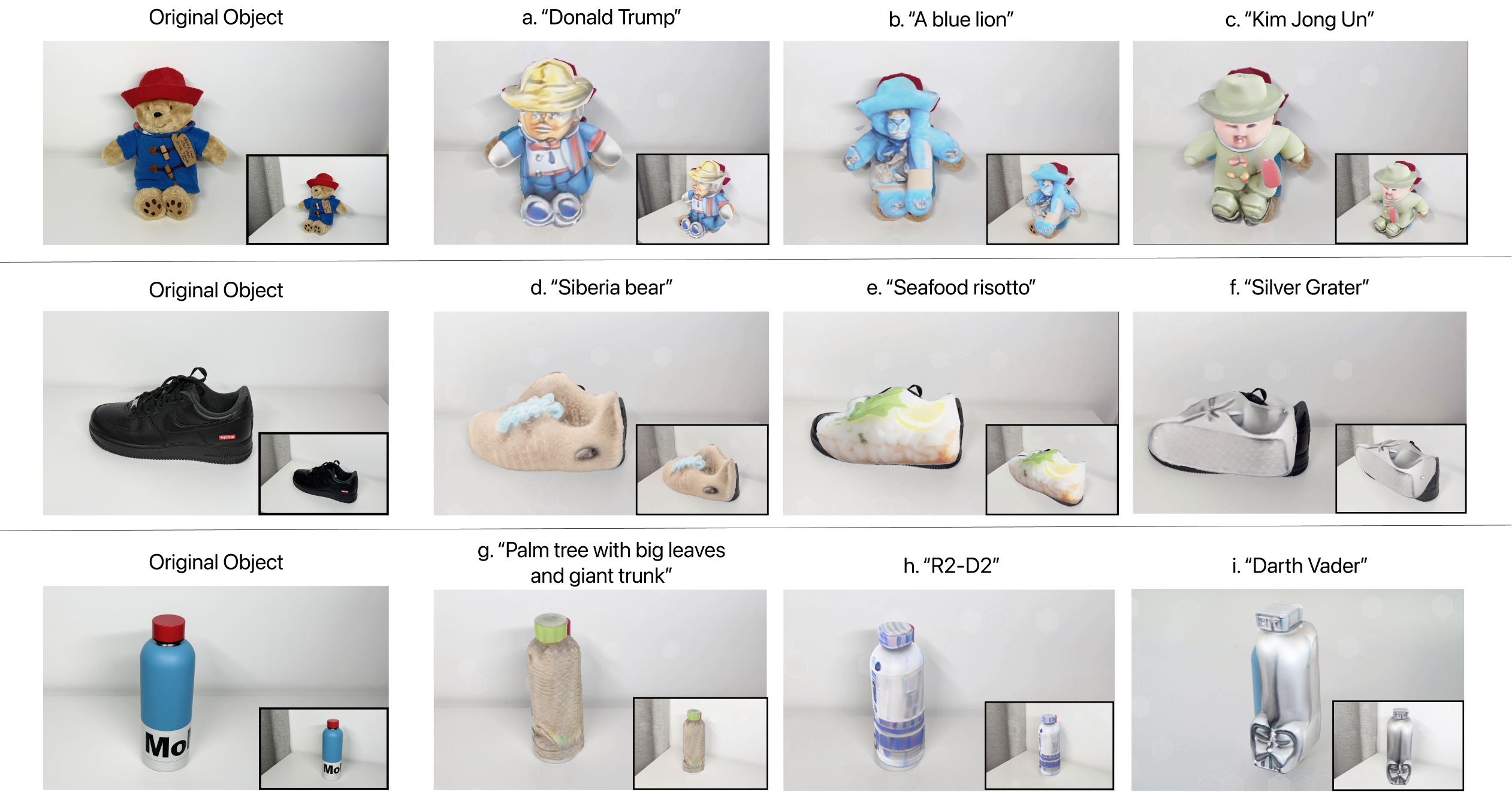}
  \caption{
  Examples of user-generated prompts and their corresponding generated results from Task 2.
  }
  \Description{ 
    This image shows a collection of nine distinct 3D models, each corresponding to a unique prompt. Here's a detailed description of each model for accessibility:
    a. "Donald Trump":
    Depicts a caricature figure wearing a blue suit, red tie, and a construction hat, with exaggerated facial features.
    b. "A blue lion":
    Features a stylized lion in a bright blue color, with a red hat and a playful pose.
    c. "Kim Jong Un":
    Shows a caricature of a figure in a green military outfit with a hat, holding a round object.
    d. "Siberia bear":
    A slipper shaped like a bear's head, complete with small ears and eyes, suggesting a cozy, thematic representation.
    e. "Seafood risotto":
    Resembles a plate of risotto with seafood, topped with a green leaf and lemon, rendered to look appetizing.
    f. "Silver Grater":
    This model is of a kitchen grater with a metallic finish, detailed to show the grating surfaces.
    g. "Palm tree with big leaves and giant trunk":
    Represents a water bottle wrapped in a palm tree texture, suggesting the trunk and leaves of a palm tree.
    h. "R2-D2":
    A slim, cylindrical model inspired by the famous "Star Wars" robot, colored in blue and white.
    i. "Darth Vader":
    A metallic flask or bottle, styled to evoke the shape and details reminiscent of Darth Vader's helmet.
  }
  \label{fig:usergeneration}
\end{figure}
\paragraph{\system{} improves user interaction and engagement.}
Our system notably improved user interaction and the quality of the haptic experience. We noticed that the participants demonstrated a smoother and more active interaction, with the ability to locate and manipulate objects with precision when using \system{} (\autoref{fig:userinteractions}). This contrasted with the baseline, where several participants missed their initial grip due to misalignment between real and virtual objects.

When asked their preferred condition, 11 of the 12 participants favored \system{}, describing it as more \textit{``realistic''} (P1, P3, P9) and \textit{``convincing''} (P2, P7, P8, P10), and the haptic experience felt more \textit{``natural''} (P2). The participants also commented on the difference between \system{} and the baseline. P8 described that moving objects with the baseline felt like \textit{``using a controller''}, while with \system{}, it felt like they were \textit{``really touching the object''}. Some participants mentioned that although the haptic feedback for the baseline was \textit{``helpful to some extent''} (P12), they doubted \textit{``how much lift it brings to the experience''} (P10). 

\paragraph{Shape alignment enhances the haptic experience.}
Participants highlighted that \system{} provided a more immersive haptic experience. Several noted that touching and feeling the object's shape and outline felt the most different between the two conditions. While interacting with the Paddington Bear, P10 remarked: \textit{``For a moment, I totally forgot it was something else there''}. P9 and P2 said that touching the bear's hat brim in \system{}, while seeing themselves touching a virtual hat in the headset, was \textit{``the most impressive moment''}. P4 shared a similar experience when touching the hand and feet of the bear toy. For the baseline, such deep experiences regarding haptic feeling were not mentioned. These results indicate that \system{}'s shape-alignment approach enhances the perceived haptic experience.

\paragraph{\system{} provided enjoyable and playful creative experiences}
\system{} not only enhanced the perceived haptic feedback but also turned the interaction into a playful and creatively engaging experience. For task two, participants overall responded very positively to the generated transformations (M=3.94, SD=0.92). Many participants described the \system{} experience as \textit{``fun''} and \textit{``interesting''}. Participant interaction during the second task was often joyful and filled with laughter during the generated transformations. Many expressed excitement while waiting for the results after inputting their prompts, and surprise upon seeing the outcomes. Many participants said they would like to try it again.

When asked \textit{``If you could take this home, where would you apply it?''} Participants offered creative answers beyond hand-held props, such as having a \textit{``Star Wars themed bedroom''}, a \textit{``cool painted bike''}, a \textit{``luxury villa''}, or turning a person into \textit{``James Bond''}. Others envisioned more practical uses for design and productivity, such as \textit{``helping generate design ideas''}, \textit{``rapid prototyping''}, or \textit{``making a pill bottle more child-friendly''}. Some participants also suggested adding a feature to recommend what to generate based on the real object. 

In addition, we observed that the quality of generated outputs, in terms of semantic prompt accuracy or precise shape alignment, did not always correlate directly with participant enjoyment. In several instances, participants responded enthusiastically despite notable deviations from conventionally anticipated outcomes. For example, when asked to generate \texttt{``Darth Vader''} (\autoref{fig:usergeneration}i), the system generated a silver column with the Darth Vader mask on the bottom instead of the head. Yet, the participant reacted positively, laughing and eagerly engaged with the object, describing it as \textit{``very funny''}. Similar reactions occurred when asked to generate a \texttt{``Seafood risotto''} (\autoref{fig:usergeneration}e), where unconventional outcomes nonetheless elicited enjoyment and engagement. These results suggest that \system{} not only served as a passive haptic tool, but also enabled potential design space for creative exploration. The wide variety of prompts chosen and participants’ reactions underscore \system{}'s significant potential as a creative medium.


\paragraph{Influence of Physical Props on Prompt Choice}
The choice of prompts varied significantly among the participants and was at times influenced by the physical props themselves. The participants gave a wide range of prompts in Task 2 (\autoref{fig:usergeneration}). Some referenced pop culture, like Star Wars, superheroes, or cartoon characters for the Paddington Bear, while others chose real-world items such as \texttt{``Palm tree with big leaves and a giant trunk''} for the bottle or \texttt{``Seafood risotto''} for the shoe. Imaginative prompts included \texttt{``Silver Grater''} for the shoe, and real figures such as \texttt{``Donald Trump''} and \texttt{``Kim Jong Un''} for the Paddington Bear. Among the props, the Paddington Bear transformations received the highest rating (M=4.17, SD=0.58), followed by the Shoe transformations (M=4.00, SD=0.95), and the water bottle transformations (M=3.67, SD=1.15).
We also observed that the participants' choice of prompts was sometimes influenced by the haptic prop. For example, when using Paddington Bear as the prop, many prompts related to animals, such as \texttt{``A blue lion''} or \texttt{``Pig''}. Similarly, with the bottle prop, prompts like \texttt{``Palm tree''} and \texttt{``R2-D2''}, which are column-shaped, were common. The variation of prompts combined with the positive response from participants showcases the highly transformative capabilities of \system{}.

\paragraph{Texture mismatch can negatively affect realism and believability.}
Although 11 of the 12 participants favored \system{} over the baseline group, one participant preferred the baseline. They found that texture mismatches, such as a hairy sensation instead of the expected pigskin for the Paddington bear, were unrealistic. They also preferred to be able to see uncovered parts of the real object, which made them feel more \textit{``secure''} and \textit{``safe''} when reaching.

Similarly, when asked what factors stopped them from trusting \system{}'s generated object, other participants also mentioned texture mismatches. For example, P7 expected a cactus prompt input to feel \textit{``prickly''} or \textit{``soft''} but the stainless steel bottle underneath felt \textit{``cold and metallic''}, with P8 noting that this kind of mismatch \textit{``reminded me this is not real''}. These comments show that in addition to shape, texture alignment also plays an important role in enhancing the believability of haptic experiences.

\paragraph{Occlusion conflicts can negatively affect immersion and spatial perception}
Conflicting occlusion cues between hands and 3D object due to tracking limitations would occasionally negatively affect the haptic experience. When the generated model was first generated, some participants perceived it as a flat layer on the screen rather than a 3D object, as the model would seem to block the hands, making it look closer to their eyes. Most participants adjusted their hand interactions accordingly to avoid confusion. However, a participant (P6, with limited VR experience, 1/5) struggled with this in later steps.

\subsection{Summary}
These findings indicate that perceived haptic realism depends not only on geometric alignment, but also on how well generated content semantically matches the physical object's affordances. Our results further suggest that effective generative passive haptics relies on the interplay between object geometry, prompt specificity, and perceptual expectations. While shape alignment significantly improves realism, mismatches in texture and semantics can break immersion, and occasionally even minor inconsistencies can remind users of the underlying physical object. At the same time, we observe that perfect fidelity is not always required: participants often remained engaged and entertained even when generated outputs deviated from expected forms. This highlights a dual role of generative passive haptics, not only as a technique for improving realism, but also as a medium for playful and creative interaction. Together, these findings point to key design considerations for generative passive haptics, where balancing physical alignment, semantic coherence, and user expectations is critical to supporting both believable and engaging experiences.
{\section{Discussion}}

We introduced \system{}, a GenAI-based MR system designed to create adaptive and on-demand passive haptic experiences. Our generative pipeline dynamically transforms arbitrary physical objects into virtual assets based on user-provided text prompts, eliminating the dependence on manual alignment, predefined mapping rules or static virtual libraries, thereby enhancing scalability. Results from our technical evaluations and user studies confirmed that the system is more effective and preferred by participants compared to a traditional static baseline. 

Our pipeline ensures real-time performance and immersiveness during user interactions. While GenAI models and pipelines have been widely used for 3D content creation, existing methods do not involve a combination of real-time user interactions, integration into a physical MR environment, and direct physical manipulation by the user with the generated result. \system{} demonstrates a viable approach for addressing the significant challenges of haptics in MR by integrating GenAI pipelines that consider the physical environment and support direct user manipulation.

Our work also contributes to understanding user behaviors, prompt patterns, tolerance to visual-tactile mismatches, and user logic behind transforming arbitrary physical objects into diverse virtual counterparts. For example, our results showed that the shape and visual design of the physical object had a significant effect on prompt choice, and that user prompts for transformation tended to follow predictable structures (\autoref{fig:userprompts}). We believe that our findings offer valuable insights that can inform designing interactive systems for customizable environmental transformation and passive haptics. 

Our generation study highlighted key factors contributing to successful outputs. Most notably, prompts that closely align with the geometry of the original physical object consistently produced better results. Interestingly, prompts that describe objects with flexible or adaptable forms, such as a \texttt{``spaceship''} or \texttt{``jewelry with silver and gold''}, were particularly effective. This result is largely due to the way that generative models interpret and transform textual input into visual content. Ensuring harmony between the user's prompt and the shape of the physical object is essential to achieve visually convincing transformations. This observation highlights an inherent trade-off between prompt fidelity and geometric alignment: prompts that closely match the physical shape maximize haptic realism, while more divergent prompts enable greater creative freedom but increase visual–tactile discrepancies. Designing systems that dynamically balance this trade-off is an important challenge for MR interaction, where both immersion and expressiveness are desirable. A promising avenue for future exploration is to investigate how the balance between ControlNet's tight alignment constraints (reducing shape mismatches) and more creative generative outcomes affects user experience and creativity. We have open-sourced our dataset\footnote{\url{https://github.com/harrywang7121/Prop-Chromeleon_Dataset}} to facilitate future comparison and research.

In our user study, participants frequently used \system{} beyond its intended purpose of effective passive haptics. Many engaged with it as a playful authoring tool, treating real-world objects as creative foundations for generating novel virtual content. Participants remained engaged even after the formal interview, frequently describing their experiences with the system as \textit{“fun”} and \textit{“interesting”}, and expressing a strong desire to keep exploring. These observations suggest that \system{} not only offers adaptive haptic feedback, but also introduces a joyful and imaginative interaction paradigm by merging GenAI with tangible, real-world references. 

Based on the results, personalizing transformations and physically interacting with familiar objects sparked delight, especially when results were humorous or unexpected. This mirrors patterns seen in broader GenAI communities~\cite{comfyui} and underscores \system{}’s unique strength in combining embodied interaction with creative exploration. As prior work has shown the transformative potential of GenAI in creative workflows~\cite{Epstein2023Art}, our findings extend this to MR, pointing toward new opportunities for interactive content creation grounded in the physical world. Our work may be particularly well-suited for applications such as early design ideation and market research, where users can quickly explore many visual variations while maintaining the same shape constraints. Similarly, our pipeline can easily be integrated with existing Tangible User Interfaces, such as \textit{Ubi-TOUCH}~\cite{Jain2023UbiTouch}, to provide tangible interaction with virtual objects that are more aligned with the physical environment yet remain within the intended theme of the virtual object and its interactions. Extending this idea, future systems could incorporate prompt recommendation mechanisms that analyze available objects in the environment and suggest transformations that better align with their geometry and affordances.

Taken together, our results highlight how the practicality of \system{} depends on the relationship between prompt semantics and prop geometry. Prompts that closely align with the physical shape enable convincing haptic experiences, as the generated content preserves both geometry and affordances. In contrast, prompts that diverge from the underlying shape remain visually plausible but increasingly rely on appearance-driven reinterpretation, where the physical object primarily supports interaction rather than faithfully representing the virtual form. This reveals a key limitation of the approach: transformations are inherently constrained by the geometry of the physical prop. As a result, many effective use cases are better suited to stylistic or surface-level modifications. This is consistent with participant behavior, where users frequently explored applications such as retexturing, stylization, or theme-based transformations that preserve the underlying structure while altering visual identity. These findings suggest that \system{} is most practical in scenarios where prompts can be grounded in existing geometry, such as design ideation, rapid prototyping, or creative exploration of variations. Understanding this relationship between prompt fidelity and geometric constraints is important for designing MR systems that balance visual expressiveness with physically grounded interaction. These findings suggest the following design considerations for generative passive haptics:

\begin{description}
    \item[Geometry over semantics:] Haptic fidelity is governed by the physical object shape, prompts should be designed to align with it for best transformation results and experience.
    \item[Stylize when shape diverges:] When prompt semantics do not align with physical geometry, favor surface-level transformations (e.g., color, texture, material, theme) that preserve structure while enabling visual reinterpretation.
    \item[Constraints as creative scaffolds:]  Make geometric limits visible to to guide user prompt selection, while embracing them for playful exploration and transformations.
\end{description}


\subsection{Safety Considerations in Object Remapping}

While \system{} enables flexible remapping of physical objects, it also introduces important safety considerations, particularly when remapping tool-like or potentially dangerous objects. For example, transforming a gun-shaped object into a harmless virtual item may lead to unsafe interactions if users misinterpret the underlying physical object or apply actions that are inappropriate in real-world contexts. Although our demonstrations are conducted in controlled environments, such scenarios highlight broader risks when deploying visual reinterpretation systems in home or public settings. To address these concerns, future systems should incorporate safety-aware transformation constraints. These may include detecting object categories with inherent risk (e.g., tools, sharp objects, or weapon-like shapes), restricting or flagging unsafe transformations, and providing contextual feedback or warnings to users. Additionally, transformation pipelines could avoid mappings that significantly alter the perceived function of an object or preserve key affordances to reduce misuse. Integrating such safeguards is critical for ensuring responsible deployment of generative MR systems.

\subsection{Future Work and Limitations}

Our current implementation relies on user-defined prompts for generation. However, the \system{} pipeline can easily be adjusted to use system-defined prompts for generation. This capability can be designed to create haptic experiences that are tailored to each user's personal environment. For example, applications and experiences can transform physical objects in the user environment to fit the narrative or needs of the system, thereby providing an immersive experience with auditory, visual, and haptic sensations. Furthermore, our current implementation strictly adheres to the shape of the physical object. However, this constraint could be dynamically relaxed in scenarios where only part of the virtual object requires haptic feedback, where the virtual shape should be prioritized, or when there is a significant mismatch between the physical object and the expected virtual shape. Introducing a dynamic weighting to shape alignment could make \system{} more flexible and robust. We believe that this avenue, together with enhancing generation with further customization such as object texture or weight for closer integration between the physical and virtual environments, is an exciting future direction for creating experiences that are customized to each individual.

In terms of creating on-demand haptics to enhance immersion, leveraging the inherent affordances of physical objects, as suggested in prior work~\cite{OpportunisticControls}, presents an exciting direction. Our user study revealed the importance of managing texture mismatches, as they can significantly impact perceived realism. Integrating advanced vision models (e.g., GPT-4o~\cite{openai2024gpt4o}) could enable estimation of physical affordances (such as texture, weight, and smell), providing richer guidance for the generation process and expanding creative possibilities rooted in varied physical properties.

Additionally, participants expressed significant interest in extending \system{} beyond handheld items to larger contexts, such as entire room transformations. Future developments could incorporate multi-object tracking techniques and integration with smart MR systems such as \emph{LLMR}~\cite{LLMR} or \emph{Remixed Reality}~\cite{RemixedReality}. Such advancements could enable users to execute complex commands like "transform this room into a spaceship cabin," by sequentially transforming multiple objects into the intended theme.

\system{} was positively received in our user study. However, our current implementation faces potential limitations related to the reliability of 3D reconstruction. While our approach maintains a practical balance between computational efficiency and accessibility, it may encounter failures under challenging conditions such as suboptimal viewpoints. Additionally, loss of fine color detail during reconstruction can result in lower-quality generative outputs. Despite these challenges, we remain optimistic about rapid advancements in this field. Future work may also focus on improving usability by exploring trade-offs between processing complexity and real-time responsiveness—for example, incorporating multi-view neural reconstruction techniques~\cite{MVR_MVD, MVR_MVLayoutNet, MVR_transformer} for more accurate physical geometry capture. We are confident that our findings, evaluation framework, and overall approach will remain relevant and adaptable as reconstruction technologies evolve.

Furthermore, a key limitation of the current system is the $\sim$20\,s end-to-end latency of the generative pipeline, largely dominated by external API calls for T2I and image-to-3D ($\sim$80\% of latency) which can be improved with more computing power. While generally tolerated in exploratory use, it introduces a noticeable waiting period that may impact continuous interaction. Future work should reduce latency through model optimization, caching, or on-device inference, and explore interaction techniques that better accommodate or mask waiting periods.

Finally, our pipeline can be expanded in several ways. We assume that the physical objects are generally solid, since changes in their shape would require regenerating the corresponding virtual object. One possible solution is to rig the virtual object to multiple anchor points, allowing it to deform dynamically as the physical object changes shape. Furthermore, enhancing occlusion cues in rendered virtual objects could significantly improve users' spatial perception and overall interaction realism. Adopting advanced spatial tracking techniques rather than traditional 2D video methods would further enhance \system{}’s experience.
\section{Conclusion}
We presented \system{}, a GenAI-driven MR system that dynamically transforms everyday physical items into shape-aligned passive haptic props based on user prompts. Through technical evaluation and user studies, we demonstrated \system{}’s effectiveness in handling diverse physical geometries and prompt types, while achieving high fidelity in both shape alignment and semantic adherence. Compared to conventional baselines, \system{} significantly enhances realism, immersion, engagement, and overall user preference. Our findings also suggest that incorporating the geometry of physical objects into prompt design leads to more coherent and believable outcomes. Furthermore, our study reveals that \system{} has the potential to serve not merely as a haptic tool, but as a creative platform that bridges GenAI with embodied MR experiences. These results highlight the broader applicability of GenAI in enabling passive haptic experiences for MR.

\begin{acks}
We would like to thank the anonymous reviewers for their feedback. 
\end{acks}

\bibliographystyle{ACM-Reference-Format}
\bibliography{Haptic}


\end{document}